\documentclass[conference]{IEEEtran}
\usepackage{cite}
\ifCLASSINFOpdf
  \usepackage[pdftex]{graphicx}
\else
\fi
\usepackage[cmex10]{mathtools}
\usepackage[cmex10]{amsmath} % MIKE: for eqref
\usepackage{url}
\usepackage{float}
\usepackage{verbatim}
\usepackage{algorithm}
\usepackage{algpseudocode}
\usepackage{graphics}
\usepackage{epsfig}
\usepackage{listings}
\usepackage{multirow}
\usepackage{amssymb}
\DeclareSymbolFont{matha}{OML}{txmi}{m}{it}
\DeclareMathSymbol{\varv}{\mathord}{matha}{118}
\newcommand{\MN}{\ifdefined\Mynote}

\begin{document}
\lstdefinestyle{myStyle}{
    basicstyle = \footnotesize\ttfamily,
    xleftmargin=0.5cm,
    frame = single,
    captionpos = b,
}

%
% paper title
% Titles are generally capitalized except for words such as a, an, and, as,
% at, but, by, for, in, nor, of, on, or, the, to and up, which are usually
% not capitalized unless they are the first or last word of the title.
% Linebreaks \\ can be used within to get better formatting as desired.
% Do not put math or special symbols in the title.
\title{Sequence Feature Extraction for Malware Family Analysis via Graph Neural Network}

% author names and affiliations
% use a multiple column layout for up to three different
% affiliations
%\author{\IEEEauthorblockN{Shun-Wen Hsiao}
%\IEEEauthorblockA{Dept. of Management Information Systems\\
%National Chengchi University\\
%Taipei, Taiwan\\
%Email: hsiaom@nccu.edu.tw}
%\and
%\IEEEauthorblockN{Your Name}
%\IEEEauthorblockA{Twentieth Century Fox\\
%Springfield, USA\\
%Email: homer@thesimpsons.com}
%\and
%\IEEEauthorblockN{Some Name\\ and Whose Name}
%\IEEEauthorblockA{Starfleet Academy\\
%San Francisco, California 96678--2391\\
%Email: email@company.com}}

% conference papers do not typically use \thanks and this command
% is locked out in conference mode. If really needed, such as for
% the acknowledgment of grants, issue a \IEEEoverridecommandlockouts
% after \documentclass

% for over three affiliations, or if they all won't fit within the width
% of the page, use this alternative format:
% 
\author{\IEEEauthorblockN{Shun-Wen Hsiao\IEEEauthorrefmark{1}
Po-Yu Chu\IEEEauthorrefmark{1}}
\IEEEauthorblockA{\IEEEauthorrefmark{1}Dept. of Management Information Systems, National Chengchi University, Taipei, Taiwan\\
\{hsiaom,109356020\}@nccu.edu.tw}
}

% use for special paper notices
%\IEEEspecialpapernotice{(Invited Paper)}

% make the title area
\maketitle

% As a general rule, do not put math, special symbols or citations
% in the abstract
\begin{abstract}
\MN
\begin{enumerate}
      \item 非數值行的不固定長度文字序列資料難處理，但是現實生活中有許多行為的profile, calls etc.
皆需被分析，尤其是資訊安全分析中有許多序列型資料難以分析
      \item 傳統分析方法：alignment, index, onehot, tfidf;
或者是新穎方式將字或句子轉成純數字：word2vec，然後再進行資安判斷但是seq分析往往忽略seq本身結構不只是一長串而已ex.
loop, repeated, noise, swap, long) alignment, 基因序列
尤其資安call中常見此現象，因為call皆為程式結構式的產出
      \item 為了更深入分析seq和他的結構，我們決定採取混合圖形表示
      那麼往後便可以將結構轉為更為容易分析的vector資料,也就是說seq內容與結構相似的資料在vector空間中會相近，便可以套用更深入的機器學習演算法分析原本難以分析的seq。
      \item problem define:
本研究要製作一個演算法可以將非固定長度的序列型文字資料轉換至向量空間表示，並且該空間可以代表此序列內容及其結構。在其後續的資安家族分類可以有良好的結果。
      \item challenge: 1. 從seq發現結構是滿困難的事  3.
而且seq資料通常很雜亂為observation可能會有noise，因此處理
      noise很重要
      \item in this study we adopt markov model, doc2vec to construct
our algorithm
      \item intro: why markov model: 紀錄seq的結構關係, doc2vec why and pros.
      \item in our experiment: classification outpeform than
benchmark. quantity: numbers, quality: graph.
      \item discussion:
      分析seq先做出圖萃取結構information再轉回vector，能更好的保留結構...
  \end{enumerate}

\fi
Malicious software (malware) causes much harm to our devices and life. We are eager to understand the malware behavior and the threat it made. Most of the record files of malware are variable length and text-based files with time stamps, such as event log data and dynamic analysis profiles. Using the time stamps, we can sort such data into sequence-based data for the following analysis. However, dealing with the text-based sequences with variable lengths is difficult. In addition, unlike natural language text data, most sequential data in information security have specific properties and structure, such as loop, repeated call, noise, etc. To deeply analyze the API call sequences with their structure, we use graphs to represent the sequences, which can further investigate the information and structure, such as the Markov model. Therefore, we design and implement an Attention Aware Graph Neural Network (AWGCN) to analyze the API call sequences. Through AWGCN, we can obtain the sequence embeddings to analyze the behavior of the malware. Moreover, the classification experiment result shows that AWGCN outperforms other classifiers in the call-like datasets, and the embedding can further improve the classic model's performance.
\end{abstract}

% no keywords % MIKE: new IEEE template has keywords

% For peer review papers, you can put extra information on the cover
% page as needed:
% \ifCLASSOPTIONpeerreview
% \begin{center} \bfseries EDICS Category: 3-BBND \end{center}
% \fi
%
% For peerreview papers, this IEEEtran command inserts a page break and
% creates the second title. It will be ignored for other modes.
\IEEEpeerreviewmaketitle

\section{Introduction}
% no \IEEEPARstart
\MN
Problem definition: We would like to investigate the information contained in the graph from sequences 'is' better than the original sequences. So that by leveraging the power of graph and GCN, we can have a better representation of sequence (embedding) for the further classification task. Also, we could extract the important events in the sequences (by using GCN attention) to specify which event is a more important feature for latter classification. The overall performance (of embedding and classification) is better than conventional ML algorithm (one-hot multiple gran), classic NN (CNN, LSTM), and cutting-edge language model (e.g., BERT). In addition, the latent space generated in the GCN can better represent the relation between sequences which matches to the evidence of case studies.
\fi

\MN
Non 
\fi

In recent years, malicious software (malware) has yielded over 200 million per year~\cite{mcafee}. Malware causes much trouble in information security because of the massive amount and rapid creation. According to the report~\cite{AV-TEST}, the number of malware has exceeded 1,339 million. Furthermore, the hackers can produce the malware through metamorphism, a technique that mutates the malware compositions but with the same purpose. Metamorphism may change the API (Application Programming Interface) calls invoked with each run of the infected program~\cite{metamorphic} and insert dummy or random calls (i.e., noise) for obfuscation. That is, malware can be massively generated within a short period and can avoid being detected. Consequently, the malware authors can produce over 12 million new malware per month on average in the last two years.

A variety of malware poses different kinds of severe problems to the user device, including smartphones, laptops, and desktops. For example, WannaCry, a branch of ransomware, caused a worldwide impact on all walks of life in 2017. WannaCry encrypts the data on the affected endpoint device and demands the ransom payment in the Bitcoin. The statistics point out that 300,000 systems in over 150 countries had been damaged because of wildly and rapidly spreading speed~\cite{WannaCry}. For another instance, Mirai is malicious software that creates botnets of IoT devices. Mirai drew the public attention after it was used in distributed denial-of-service (DDoS) attack against the website of Kreb on Security, which covers computer security and cybercrime~\cite{Mirai}. The DDoS attack launched through Mirai caused the crash on high pageview websites, including GitHub, Twitter, Reddit, etc. From the example stated above, we realize how desperate damage can be caused by the malware. 

However, the current malware analysis approaches highly rely on malware analysts' domain knowledge and standard operation process to identify the intention of malware. In contrast, the metamorphism technique hugely enhances the difficulty of malware analysis. The massive amount of malware and the sophisticated call sequence invoked by malware makes the analysis difficult. The complicated and innumerable tasks overwhelm the analysts. Consequently, an automatic and timely analysis method is indispensable to achieve proper defense and lessen the infliction of loss on the computer devices.

%各補cite static&dynamic
To analyze the behavior of malware, static and dynamic analysis methods are usually applied. Static analysis method extracts program behavior form collected executable file, such as Windows portable executable (PE) file or Linux executable and linkable format (ELF) file~\cite{Staticanalysis}. Dynamic analysis method executes malware in a sandbox and records the interaction between malware and operation system~\cite{Dynamicanalysis}. These static and dynamic malware profile is usually unstructured, text-based, temporal, and/or variable-length that makes it more difficult to analyze. In this paper, we focus on dynamic analysis profile that contains a sequence of call invocation executed by malware.
% It is challenging to deal with variable length and non-numeric sequential text data. However, many behavior data, such as profiles and calls are text-based sequential data with variable lengths that need to be analyzed, especially in the information security domain.

It is challenging to deal with variable length and non-numeric sequential text data.
In the past, several methods are used when analyzing sequential text data, such as sequence alignment~\cite{Seqalignment}, bag-of-words, or word2vec~\cite{Word2vec}. These methods transform the text in the profile into a numerical vector but their transformation do not consider the structure of the text sequence. In such way, a malware author may construct a malware through the permutation of call sequence and the insertion of noise fragments~\cite{MS} to avoid being detected such methods. Namely, unlike natural language, the call sequence may contain certain structure, such as loop, repeated call, and swap structure generated by the malware code. From this point of view, when transforming the text into numerical vector, we further consider to include the structure of the text to better represent a malware profile. This study investigates using a graph, such as finite state machine or Markov model, to represent the structure of the call sequence. We then embed the structure and text together to represent a malware profile. We believe that such representation of profiles can help the downstream security applications, for example, malware family classification, malware characteristics extraction, malware behavior clustering. 
% cite family classification ...
% the representation and state awareness extracted from the graph outperformed the sequence type.
% The proposed method can well represent the text and the structure used by a vector, and  conventional text embedding method.

In this study, we would like to develop a novel graphic neural network to automatically embed variable-length, sequential, text-based data to a proper numerical vector with preserving its hidden information in the sequence structure. In addition, this method can highlight the most important sub-sequence in the profile when performing the downstream task. In this way, we can visualize the importance of individual call in the profile. We anticipate the embedding methods should outperform conventional text embedding methods~\cite{DetectMalware, Malwaredection}.

% we need a design to investigate the information and structure hidden in API call sequences. Hence, our study’s goal is to design an appropriate approach suitable for the call-like data, which can preserve the information of the API call sequences further and highlight the state awareness of the graph.

% challenges and and solution
To achieve our goal, the following issues should be solved:
(1) how to represent a structure of a sequence and text information while considering the sequence is variable-length and contains noises;
(2) how to embed the graph structure, node (call features and information), and edge (call transition probability and information) into a vector;
(3) how to identify the importance of each individual call in the sequences when performing downstream task.
% (4) Identifying whether the attention weight is correct or not is an arduous task. Since the judgment of attention position have no criteria.

For the first issue, we design and implement a Graph Generation Module that adopt the Markov model to transform each sequence to a corresponding graph with the transition probabilities learned from the data and the information of calls. By our design, we can preserve the structure information hidden in the original call sequences and handle the imbalance length of sequences.
% We first proposed the Graph Generation module to conserve further information remaining in the original API call sequences. Meanwhile, the Graph Generation module can handle the imbalance length of sequences.
For the second issue, we then feed the graphs to the customized graph convolution neural network (GCN). While training the neural network with each graph, the node information between the consecutive calls in the graph will be exchanged bidirectionally regarding to its transition probability. The last layer of the customized GCN is designed as the representation of the input graph.
For the last issue, we design a novel attention structure that can specify the importance of each individual call from the training data directly and automatically. This structure helps us to reveal the importance of input feature and transfer the attended features to the next convolution layers.
At the end, we can obtain the importance of each call of the sequence with the help of graphic structure.

% In spite of the difficulty listed above,  we still apply ourselves to dig out the solution to the analysis process. To solve the problem we faced, we proposed \textbf{Attention Aware GCN (AWGCN)}, which can mainly be split into two modules. The other is our graph neural network model, including attention, graph convolution, and dense layers. This model helps us extract the feature precisely and obtain attention to the graph. 

The contributions of this study are listed as follows.
\begin{itemize}
\item We develop a graph-based approach to dealing with the embedding issue for the variable-length, sequential, text-based sequences data whose structure contains loops, repeated events, shifted sub-sequence, etc. We anticipate that adopting graph representation is more appropriate than using NLP approach to analyze such sequential data. 
\item We develop a GCN-based neural network to transform the graphs to numerical vectors that can well perform the representation of the information contained in the graphs for the latter downstream task, such as malware family classification.
\item  We develop a novel attention structure directly on the input text (node) and in the convolution structure to capture the attention that can help us reveal the crucial sub-sequence of the input.
\item We use the real-world malware datasets to demonstrate the effectiveness of the proposed system compared with the several embedding models and NN-based models. The proposed Attention Aware Graph Convolution Network (AWGCN) can better represent the sequence data and outperform them on the downstream malware family classification tasks. On four datasets with more than 6,000+ malware samples, we have average 97.63\% f1-score which is around 2\% of improvement better than the 2-nd best models.

%the entire process from the input sequences to the output, including graph visualization, malware family classification, and graph attention. Applying our solution can help the analysis process of call-like sequential data.
\end{itemize}

The rest of the paper is organized as follows. In Section \ref{sec:related}, we review some background and related works. Section \ref{sec:proposed system} is the proposed design of AWGCN. The evaluations of AWGCN are demonstrated in Section \ref{sec:evaluation}. In Section \ref{sec:discussion}, we discuss the some insights and future works. At last, Section \ref{sec:conclusion} contains some concluding remarks.

% last edition 
% Prospect of future improvement
% remaining section description

\section{Related Work}\label{sec:related}

\subsection{Text Embedding Algorithm}
For the computation convenience, usually non-numeric data (e.g., text and categorical data) are converted to a unified numeric format before performing the downstream tasks. In this section, we briefly introduce several approaches to deal with text data.
\subsubsection{Bag-of-words and One-hot}
Bag-of-words uses word occurrence frequency in the sequence to convert the original text-based sequence. One-hot converts a word into a vector with only one digit being one; others are zeros. The disadvantage of such approaches is the curse of dimensionality if lots of unique words are used.
\subsubsection{Word2Vec~\cite{Word2vec}}
There are two training models in Word2vec; one is skip-gram, and another is CBOW. They are both self-supervised learning method that uses sentences to train the word vector. Skip-gram inputs a word to predict the context; on the other hand, CBOW inputs the context to predict the blanking word, like a cloze task. Because Word2vec generate a single word embedding representation for each word in the corpus~\cite{Bertwiki}, it can not deal with polysemy words.
\subsubsection{Doc2Vec~\cite{Doc2vec}}
To overcome the shortcomings of bag-of-words that lack of information of word ordering and sentence structure. Le and Mikolov proposed Paragraph Vector, adding paragraph identification into the model. This mechanism takes the sentence position in the paragraph into account. %Paragraph identification is seen as the feature of the sentence.
\subsubsection{RNN}
Recurrent Neural Network (RNN) uses the previous hidden state to predict the next state, and is widely used to analyze sequential data. Hochreiter and Schmidhuber developed long short-term memory (LSTM)~\cite{LSTM} networks to improve original RNN and performed well in multiple application domains.

\subsubsection{Transformer~\cite{transformer}}
A transformer is a deep learning model consisting of self-attention mechanism and feed forward neural network. The structure of transformer is the stack of encoders and decoders. The design of transformer deprecates the mechanism of RNN, namely, the prediction of input can learn from any part of the original sequence. Transformer are increasingly the model of choice for NLP problems, replacing RNN models such as long short-term memory (LSTM). This also led to the development of pretrained systems such as BERT~\cite{transformerwiki,bert}.

\subsection{Graph Neural Network}
Convolution layer have been widely used for extracting higher-level representation from image data for latter analysis, for example, extracting facial features from a pixel image for emotion detection. Convolution Neural Network (CNN) has been proven well-performed in many computer vision domains. However, analyzing image data by CNN relies on the fixed input order and neighbor pixels to correctly extract higher-level features ~\cite{DGCNN}. 

Nowadays, the concept of convolution has been extended to other data type, such as graph who has no fixed ordering and has more complex neighboring relations. To analyze a graph, e.g., knowledge graphs and molecular structure, convolutional neural networks have been generalized on the graph domain ~\cite{cnnongraph}.

According to the taxonomy in~\cite{ComSurvey}, graph neural networks can be categorized into recurrent graph neural networks (RecGNNs), convolutional graph neural networks (ConvGNNs), graph autoencoders (GAEs), and spatial-temporal graph neural networks (STGNNs). Moreover, GNNs can cope with three kinds of graph analytics tasks with different graph structures as input, including node-level, edge-level, and graph-level tasks. Each category can be used to deal with different main tasks as well. To classify the graph with a distinct structure, we need to learn the whole graph representation. Therefore, we focus on convolutional graph neural networks in the research, which is further close to our primary goal.

Given a graph \textit{G = (V, e)}, which consist of \(|\)\textit{V}\(|\)  vertices and \(|\)\textit{e}\(|\) edges, we can construct the adjacency matrix \textit{A}, which is a \(|\)\textit{V}\(|\) x \(|\)\textit{V}\(|\) matrix with \textit{(i, j)} entry equaling to 1 if there is an edge connecting vertex i and j and 0 otherwise. The original graph convolutional network (GCN)~\cite{GCN} proposed the following layer-wise propagation rule:
\begin{equation}
H^{(l+1)} = \sigma(\hat{A}H^{(l)}W^{(l)})
\end{equation}
where $\hat{A} = \Tilde{D}^{-\frac{1}{2}}\Tilde{A}\Tilde{D}^{-\frac{1}{2}}$, and $\Tilde{A} = A + I_N $ is the adjacency matrix \textit{A} with added self-connection. $I_N$ is the identity matrix, $\Tilde{D}$ is the degree matrix and $W$ is the trainable weight matrix. $\sigma$ denotes an activation function, such as ReLU, Softmax. $H^{(l)}$ is the matrix of activation in the $l^{th}$ layer, and $H^0 = X$, which is the feature matrix of the vertices in the graph.

\begin{figure}[t]
\centerline{\includegraphics[trim = {0mm 0 0 0}, clip, width=0.48\textwidth]{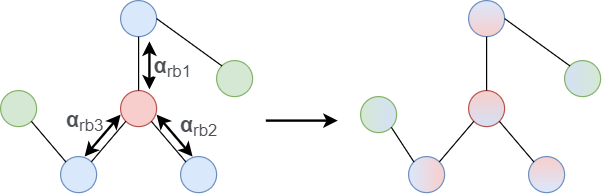}}
% <left> <lower> <right> <upper>
\caption{The visualization of information propagation process in graph convolution layers.}
\label{fig:GCN_coloredit}
\end{figure}

We can visualize the information propagation in the graph convolution layer in Fig. \ref{fig:GCN_coloredit}. GCN~\cite{GCN} assigns $\alpha$ depending on each node's degree. The nodes can acquire information of their neighbor nodes. The figure shows that the red node receives the information from three blue nodes, and all the other nodes receive information from their neighbors as well.

Owing to the success of GCN, a great deal of research proposed their own propagation rule to improve the performance further. For example, graph attention networks (GAT) ~\cite{GAT} assume the neighbors' contribution is imbalanced due to the importance being different to the central vertex. Thus, GAT implements the attention mechanism to learn the relative weights between two connected vertices before the message propagation. Graph sample and aggregate (GraphSAGE) ~\cite{GraphSage} proposed a trick to improve the efficiency of taking the total size of the neighboring vertices. GraphSAGE adopts sampling to obtain a fixed size of neighbors for each vertex and apply an aggregation function to the embedding.

In previous work, graph convolutional network are applied to graph tasks such as citation networks, social networks, biochemical graphs, knowledge graphs, and others. Furthermore, graph convolutional network are also used on text data like fake news prediction. In fake news predictions, the graph data are used to describe the information between the news authors. Namely, the graph data is additional information in fake news data. In our work, the text data (call) is regarded as a node; that is, we use graph convolutional network directly on API call data. 

Our proposed model used the graph convolutional layers to propagate the information between the vertices for the following downstream tasks. Further detailed design of our model is described in Section~\ref{sec:proposed system}.

\subsection{API Call Sequence}
API stands for application programming interface, a type of software interface that specifies how clients should interact with software components~\cite{APIDesign}. Nowadays, many service providers offer their API to allow users to extract or transport the data~\cite{Fostering}, such as e-commerce website.

When users want to interact with the system, for example, log on to the app or search the question via a browser, users have made an API call. Definitely, not only interact with browsers but also access to the system resources need the help with API. For instance, user applications cannot access hardware and system resources directly in Windows operating system. Nonetheless, they can rely on the interfaces provided by dynamic-link libraries~\cite{WinMalware}.

As stated above, the malware authors can create the user program and use the API calls to perform malicious actions. The Windows API function calls fall under various functional levels such as network resources and libraries~\cite{ZerodayMal}. Hence, the API call sequence can reflect the behavior of the execution process, whether the process is malicious or benign. In other words, we can extract information from the API call sequence to represent the process; theoretically, it should be precise for the following downstream tasks. 

\begin{figure}[t]
\centerline{\includegraphics[trim = {0mm 0 0 0}, clip, width=0.48\textwidth]{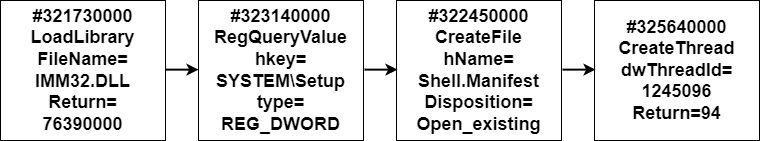}}
% <left> <lower> <right> <upper>
\caption{A partial sample of Windows API call sequence.}
\label{fig:API_call_seq}
\end{figure}

Figure~\ref{fig:API_call_seq} is a partial example of the collected malware profile. Besides the API name, an API call invocation has time information so that we can link them as a sequence as a whole. In addition, most of the dynamic analysis tool also output certain extra information, such as call parameters and return values.

\subsection{Markov Model}
A Markov model is a stochastic model to model the changing processing between each state. Markov chain is a specialized Markov model, which describes a sequence of possible events transition with their probability. With the Markov property, the probability of moving to the next state depends only on the present state and not on the previous states~\cite{Markovwiki}. The formal definition of Markov chain in mathematics is shown in Eq.~\ref{markov}

\begin{equation}
\begin{split}
Pr(\textit{$X_{n+1}$} = x | \textit{$X_1$} = \textit{$x_1$}, \textit{$X_2$} = \textit{$x_2$}, ... , \textit{$X_n$} = \textit{$x_n$})\\ = Pr(\textit{$X_{n+1}$} = x | \textit{$X_{n+1}$} = x )
\end{split}
\label{markov}
\end{equation}

\begin{figure}[t]
\centerline{\includegraphics[trim = {10mm 10mm 10mm 10mm}, clip, width=0.4\textwidth]{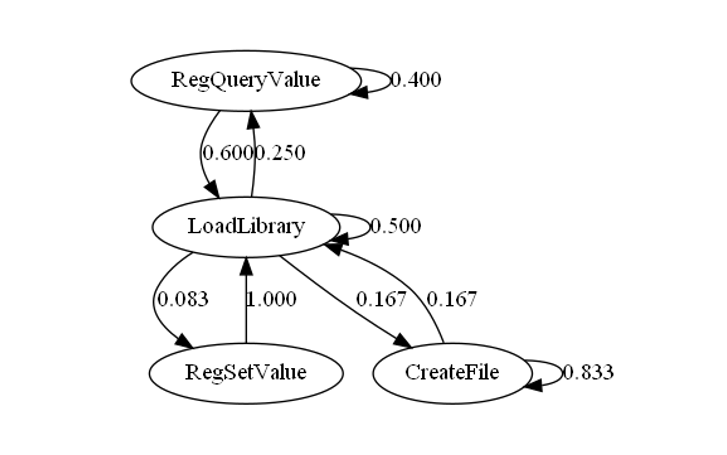}}
% <left> <lower> <right> <upper>
\caption{A real example of API call sequence Markov model.}
\label{fig:MarkovChain}
\end{figure}

Our research uses the Markov chain to represent the transition between the API call (as shown in Fig.~\ref{fig:MarkovChain}), each node represent an unique call in the sequence) and transform the original sequential data into graph format data. Figure~\ref{fig:MarkovChain} demonstrates the Markov chain of a real-world malware sample used in our experiment. This demonstrated malware belongs to Vobfus malware family, and it only invoked four unique Win32 APIs, i.e., RegQueryValue, LoadLibrary, RegSetValue and CreateFile. It has 26 calls in total. (Due to the page limitation, we select a relativity short sample for demonstration. On the average, the malware samples in our different datasets have 50 -- 364 calls in a profile.)

\section{Proposed System}\label{sec:proposed system}

\begin{figure*}[htbp]
\centerline{\includegraphics[trim = {0mm 0mm 0 0}, clip, width=\textwidth]{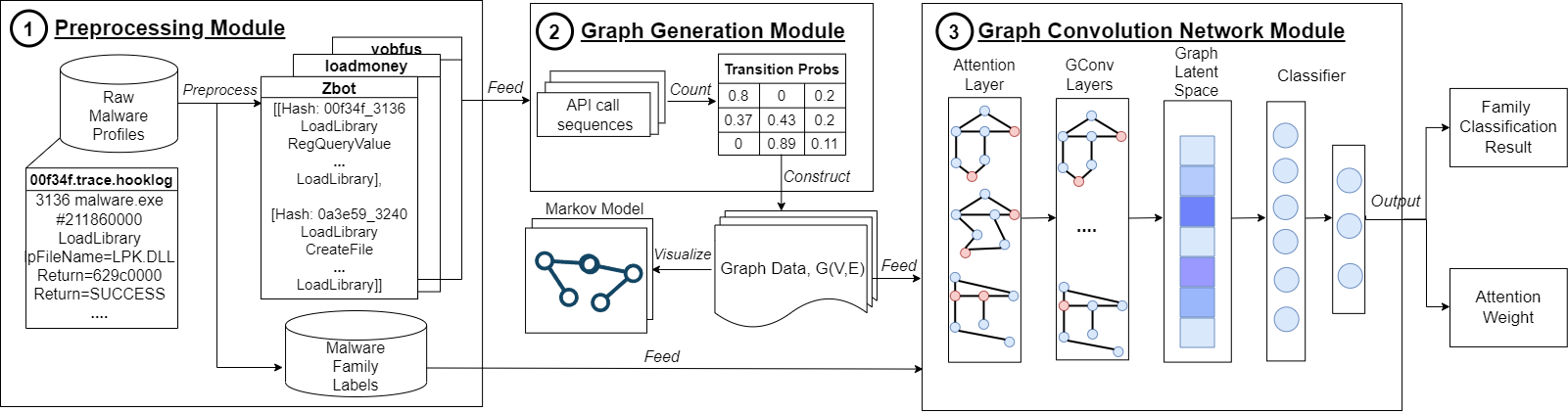}} % trim{l, b, r, t}
% <left> <lower> <right> <upper>
\caption{The overall process of our purposed system.}
\label{fig:Ch3_fig}
\end{figure*}

\begin{figure*}[htbp]
\centerline{\includegraphics[trim = {0mm 0 0 0}, clip, width=\textwidth]{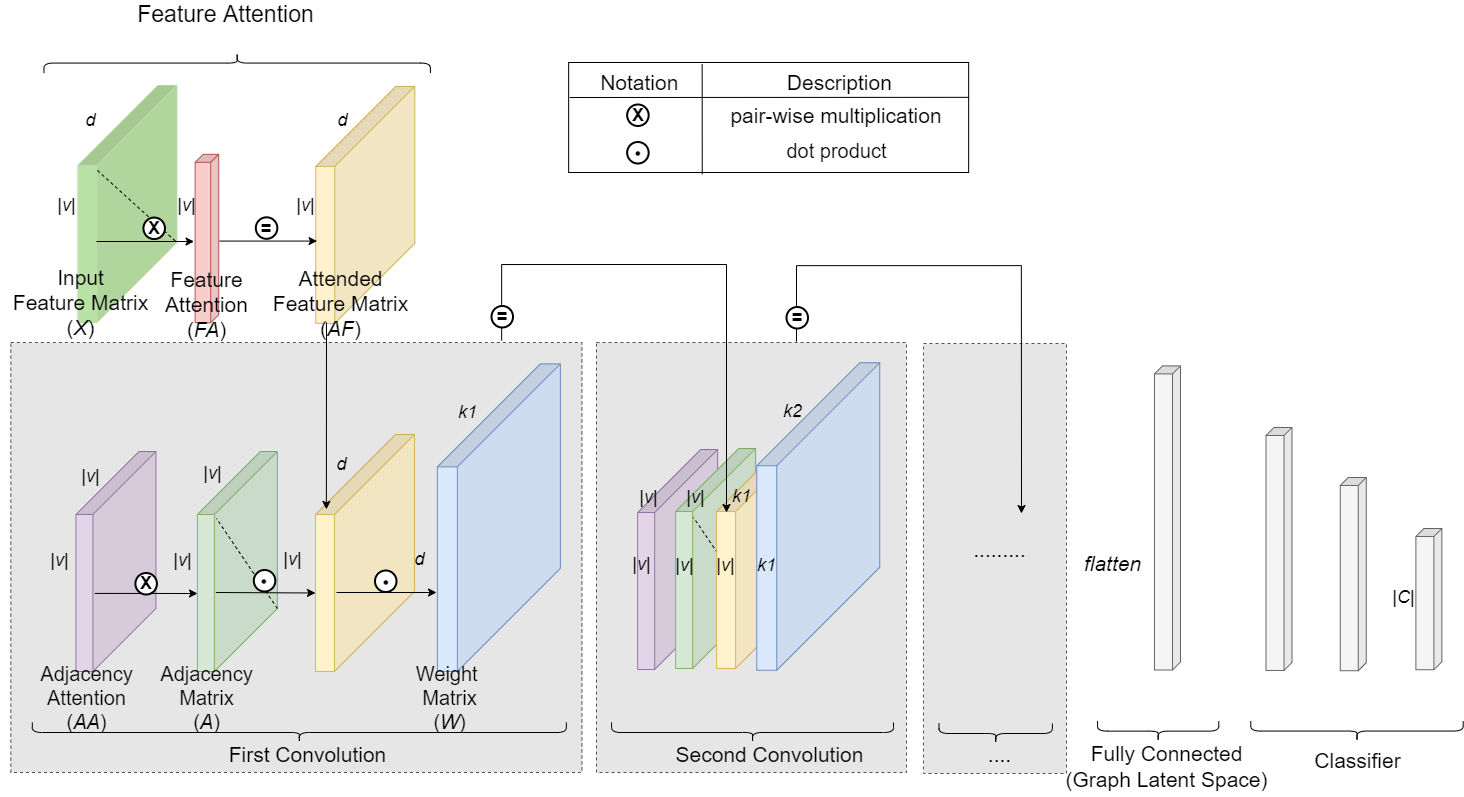}}
% <left> <lower> <right> <upper>
\caption{Graph Convolutional Network architecture}
\label{fig:model_arch}
\end{figure*}

\subsection{Overview}
Figure \ref{fig:Ch3_fig} is the overview of the proposed system that can input the raw malware profiles (in terms of API/system call sequences) for graph representation, embedding, graph attention and downstream malware family classification.

The system has three main modules: Preprocessing, Graph Generation Module and Graph Convolution Network.
The Preprocessing module parse the raw data (i.e., malware profile) and malware family labels to construct the call sequences of each malware family.
The Graph Generation Module accepts all sequences and generate a Markov model graph with transition probability for each sequence. Vanilla Markov model graphs can be plotted in this stage.
The Graph Convolution Network accepts the training data (i.e., the graphs) and the training labels (i.e., corresponding malware family labels) to perform graph embedding by AWGCN. The Graph Latent Space represents the original call sequence with the graph structure. The neighbor vectors in the Graph Latent Space indicates that their original sequences use similar calls and similar structure.
At the end, a malware family classifier can classify a profile into the corresponding class. We will use 20\% of the data for evaluation. In addition, the trained attention weights can be used to specify the importance of each call in the Markov model graph.

The detail design of each module is described as follows. The notations used in this paper are listed in Table~\ref{notation}.

%In order to properly extract features from the input sequence, we need to transform the original sequence into the graph to fit our following neural network input. Hence there are two main parts in our designed system.

% First of all, we implement a Python script to retrieve the useful information from the original hooked log file, including malware's hash value and API call sequence. In the second step, we transform the API call sequence into the Markov model. Therefore, we can obtain the transformation probability of each call, and visualize the API call sequence into a graph format. In the third step, we input the whole graph set with ground truth labels into our proposed graph neural network and extract sequence features from the graph.
\begin{table}[t]
\caption{The table of Notation}
\begin{center}

\begin{tabular}{|l|l|}
\hline
\textbf{Notation} & \textbf{Description}\\
\hline
$P$ & Profile set\\
\hline
$S$ & API call sequences set\\
\hline
$|P|$, $n$ & The number of profile files in $P$\\
\hline
$|S|$ & The number of API call sequences in $S$\\
\hline
$p_i$ & The $i$-th profile, $i = 1 \sim n$ \\
\hline
$p_i$.\textit{hash} & The hash value of $p_i$ \\
\hline
$p_i$.\textit{label} & The malware family label of $p_i$\\
\hline
$s_i$ & The API call sequence for $p_i$, $i = 1 \sim |S|$ \\
\hline
$c_j^i$ & The $j$-th call of $s_i$, j = $ 1 \sim |s_i|$\\
\hline
$c_j^i$.\textit{name} & The name of call $c_j^i$\\
\hline
$c_j^i$.\textit{pars} & The parameters of call $c_j^i$\\
\hline
\hline
$g$ & The graph of a $s$ and a $p$, $g$=($V$, $E$) \\
\hline
$G$ & The graph set of all $g$\\
\hline
$V$ & The vertex set of a graph $g$\\
\hline
$E$ & The edge set of a graph $g$\\
\hline
$v$ & A vertex in $V$\\
\hline
$e(v_a, v_b)$ & An edge from vertex $v_a$ to vertex $v_b$, $v_a$ and $v_b \in V$ \\
\hline
$v.edges$ & The set of all edges starting from $v$\\
\hline
$e.weight$ & The weight of edge $e$\\
\hline
$g.X$ & The input feature matrix of graph $g$\\
\hline
$g.A$ & The adjacency matrix of graph $g$\\
\hline
$g.A_{attr}$ &  The transition probability of the edges.\\
\hline
$W$ & Weight matrix\\
\hline
$H_l$ & The convolved feature matrix at $l$ layer.\\
\hline
$g.FA$ & The feature attention matrix of $g$ \\
\hline
$g.AA$ & The adjacency attention matrix of $g$\\
\hline
$g.AF$ & The attended feature matrix of $g$, $g.X \times g.FA = g.AF$\\
\hline
$\varv$ & The set of vertex in entire dataset.\\
\hline
$OH()$ & Do one-hot encoding to the input.\\
\hline
\end{tabular}
\label{notation}
\end{center}
\vspace{-3mm} % reduce blank between table and text
\end{table}

\subsection{Preprocessing Module}
The collected datasets have their own format to represent the call sequences, so preprocessing is needed. The Preprocessing Module will parse the profiles and select necessary information (e.g., malware hash value, timestamp, call name, call parameters, call return values) to construct the call sequences and identify its malware family label. The psudo-code of this module is shown in Fig.~\ref{fig:Data_processing_pseudo}. For each profile, $p$, in the collected profile pool, $P$ (having $n$ profiles), a sequence $s$ contains all the information of this profile, including the profile hash value, malware family label, and a sequence of call name and its parameters. In our implementation, the sequences are stored in a customized Python data structure.

%In the first part of our design, \textcircled{1} in Fig.~\ref{fig:Ch3_fig}, we retrieve the needed information from our original log file hooked from the malware process. In our implementation, we only keep the API call name at first. Since less information remains from initial data, we can test the ability of our proposed process and compare it to other models as well. If we want to improve the performance, we can add more information from the hooked log file, such as dynamic link library and return value, as well as in another field's data.

\begin{figure}[t]
\begin{algorithm}[H]
  \caption{Data Preprocessing}
  \hspace*{\algorithmicindent} \textbf{Input:} A set of profile $P$.\\
  \hspace*{\algorithmicindent} \textbf{Output:} A set of sequence $S$.
  \begin{algorithmic}[1]
    \State $S =$ EmptySet()
    \For{$i \gets 1$ to $i= |P|$}
        \State $s_i.hash = p_i.hash$
        \State $s_i.label = p_i.label$
        \State $s_i.seq = $ EmptyList()
        \For{$c \in p_i$}
            \State append ($c.name$, $c.pars$) to $s_i.seq$
        \EndFor
        \State add $s_i$ to $S$
    \EndFor
  \label{code:Data_preprocessing_pseudo}
  \end{algorithmic}
\end{algorithm}
\caption{The pseudo code of data preprocessing}
\label{fig:Data_processing_pseudo}
\end{figure}

\subsection{Graph Generation Module}
\begin{comment}
% 為什麼選用Markov model & 特色 
% 保留前後關係，不只是使用哪個字
% MM 可以保留短期的前後關係，還有量化的機率，
% 若要長期可以考率k-gram
% 機率還可以用來和 下一個模組的 ＧＣＮ 結合，用作資訊量的轉移

% 吃algo 1 seq
% 計算 trans matrix
% 生G(VE) + 數值
% 請看 algo2
% given S, generate G
% for s in S, 
% g .....

% G-> onehot; for all g=(E,V) in |G|, v in V, e in E,
% OH = union(e)
% matrix := big_V x big_v $big_V$
\MakeUppercase{text}
$\MakeUppercase{\varv}$
\end{comment}
We design a $k$-gram mechanism, which helps us to better portray the feature when transforming the sequences. We use Markov model to depict our original API call sequences because it contains a transition probability table that can describe the transition of neighboring calls.
The characteristic of API call sequence is the transition between each call because API call sequence represents the continuous occurrence of calls. This characteristic is similar to the essence of Markov model; thus, we adopt Markov model to transform the sequences. Moreover, Markov model can preserve the short-term relation of the calls and quantized transition probabilities, which can be used in the next neural network module to enrich the information of input sequences. In this form, transition probabilities can be seen as the transfer of information in the following graph neural network. In addition, if we want to preserve long-term relations, we can adjust $k$-gram mechanism to fulfill.
The psudo-code of this module is shown in Fig.~\ref{fig:GGM_pseudo}. We traverse each sequence, $s \in S$, to record the transition between the calls and add to $g.A$. In the meanwhile, we record the edge occurrence in $e.weight$ and divide by sum of $v.edge.weight$ to calculate the probability of each edge. To feed the graph data into the following graph neural network, we use one-hot encoding to transform the calls into vector format data $g.X$. We use one-hot because it is a primary encoding method that can show our following graph neural network capability. At the end, we can also visualize the $g$, such as Figure.~\ref{fig:MarkovChain}

\begin{figure}[t]
\begin{algorithm}[H]
  \caption{Graph Generation Module}
  \hspace*{\algorithmicindent} \textbf{Input:} A set of sequence $S$.\\
  \hspace*{\algorithmicindent} \textbf{Output:} $G$.
  \begin{algorithmic}[1]
    \State $G =$ EmptySet()
    \For{$s_i \in \textit{S}$}
        \State $g_i.X = OH(\varv)$
        \State $g_i.A =$ EmptyList()
        \State $g_i.A_{attr} =$ EmptyList()
        \For{$c.name \in s_i$}
            \If{ $e(c_j^i.name, c_{j+1}^i.name)$  not in $g_i.A$}
                \State append $e(c_j^i.name, c_{j+1}^i.name)$ to $g_i.A$
                \State $e(c_j^i.name, c_{j+1}^i.name).weight$ += 1
            \EndIf
        \EndFor
        \State add $g_i$ to $G$
    \EndFor
    
    \For{$v$ in $g_i \in G$}
        \State freq $=$ sum up $v_i.edge.weight$
        \State probs $=$Divide $e.weight \in v_i$ by freq
        \State append probs to $g_i.A_{attr}$
    \EndFor
    
    \label{code:GGM_pseudo}
  \end{algorithmic}
\end{algorithm}
\caption{The pseudo code of Graph Generation Module}
\label{fig:GGM_pseudo}
\end{figure}

\begin{comment}
\begin{table}[htbp]
\setlength{\tabcolsep}{2pt}
\caption{Example of transition matrix}
\begin{center}
\begin{tabular}{|c|c|c|c|}
\hline
&\textbf{LoadLibrary} & \textbf{RegQueryValue} &\textbf{CreateFile}\\
\hline
\textbf{LoadLibrary}& 
0.5455 & 0.3636 & 0.0909\\
\hline
\textbf{RegQueryValue} & 0.5000 & 0.5000 & 0.0000\\
\hline
\textbf{CreateFile} & 0.5000 & 0.0000 & 0.5000 \\
\hline
\end{tabular}
\label{trans_matrix}
\end{center}
\vspace{-3mm} % reduce blank between table and text
\end{table}
\end{comment}

\subsection{Graph Convolution Network Module}
In the beginning of our GCN model architecture, we design a Feature Attention ($FA$) layer. We want to capture the important part of the input graphs through this layer.
Next, we use \textit{k}-layers of graph convolution layers to extract the higher-level node representations of the whole graph. The graph latent space is the embedding of original $s$ in malware profile with graph structure. In the last of this model, we apply a classifier, which map the extracted feature to the label space, softmax activation function for mutil-label classification, and sigmoid for binary classification.

%\subsection{GCN Architecture}
We take a graph $G$ and the label $P.label$ as input. In Figure.~\ref{fig:model_arch}, all the cuboid with diagonal dashed line is fixed, others are trainable. We use a graph as input to explain the process in Graph Convolution Network Module. The input feature matrix in Figure.~\ref{fig:model_arch}, means the graph contains $|V|$ vertices, and each node has a $d$ dimensional feature. We then multiply the $X$ with proposed trainable Feature Attention layer ($FA$) to capture the significance of each node. As shown in Figure.~\ref{fig:model_arch} feature attention part, we get the Attended Feature Matrix ($AF$).

After that, we pass $|V|$x$d$ $AF$ to designed three times graph convolution to exchange the information between the vertices. Before the dot product begins, we multiply the adjacency matrix with proposed trainable Adjacency Attention. Hence, we can further capture the connection importance of the graph. As mentioned in~\ref{sec:related}, if vertices are connected, the corresponding adjacency matrix digit is marked as 1. Thus, we can achieve the information exchange in the $AF$ through the dot product with attention attended adjacency matrix. Furthermore, we use the transition probabilities, $g.A_{attr}$, calculate from previous module to decide the ratio of the transfer between connected vertices. In the last of a graph convolution, we use a weight matrix whose dimension is $d$x$k1$ to transform the feature dimension of nodes. Through this weight matrix, we can project the original $d$ dimension feature to a larger $k1$, which means we can adopt more information from the transferred feature matrix, this can help us learn more from the original input graph as well. 

After the first graph convolution, we obtain a new $AF$ with dimension $|v|$x$k1$. We take this attended feature matrix as input to the second convolution process. Same as the first convolution, we can adjust $k2$ to preserve more information which may help the downstream classification task. After the convolution process, we apply fully connected layer, dropout layer and use softmax to classify the multi-label task. 

The pseudo code of Graph Convolution Network is shown in Fig.~\ref{fig:GCN_pseudo}. We input the graphs $G$ with the label $P.label$ to proposed Graph Convolution Network. In the beginning, we split the whole dataset into training and testing with 80:20 ratio. We input our training graphs into the GCN, and provide the $g.X$, $g.A$ and $g.A_{attr}$ to train the GCN. In each epoch, we attend the $g.X$, convolve the $AF$ three times and use the classifier to evaluate the correctness. Then do backpropagation and update the weight for the training in next epoch. The output of this module, besides the classification result, we also obtain the graph embedding and  weight of $W$, $FA$ and $AA$. The embedding and weight can help us cluster the original sequences and specify the important part of the graphs.

\begin{figure}[t]
\begin{algorithm}[H]
  \caption{Graph Convolution Network}
  \hspace*{\algorithmicindent} \textbf{Input:} $G$, $P.label$.\\
  \hspace*{\algorithmicindent} \textbf{Output:} Trained weight of $W$, $g.FA$, and $g.AA$.
  \begin{algorithmic}[1]
    \For{$g \in \textit{G}$}
        \State Training set $\gets$ $g_i$ where i$\geq r*|S|$ \Comment{$r = ratio$}
        \State Testing set $\gets$ $g_i$ where $r*|S|<$i$\leq r*|S|$
    \EndFor
    \For{\textit{epoch}}
        \State \textit{AF} $\gets$\textit{g.X} $\times$ \textit{FA}
        \State \textit{$H_{l+1}$} $\gets$ \textit{AA} $\times$ \textit{g.A} $\cdot$ \textit{$H_l$} $\cdot$ \textit{$W_l$} \Comment{$H_0 = AF$}
        \State Bcakpropagate and update weight
    \EndFor
  \label{code:GCN_pseudo}
  \end{algorithmic}
\end{algorithm}
\caption{The pseudo code of Graph Convolution Network}
\label{fig:GCN_pseudo}
\end{figure}

\section{Evaluation}\label{sec:evaluation}

\subsection{Data Set}
As shown in Table~\ref{dataset}, five datasets are used in our experiments. SynData is used for proof-of-concept to make sure the synthetic special calls can be identified. RanSyn is sophisticated synthetic data to imitate the invocation of API calls. WinMal and Oliverira are malware profiles that record the Windows API calls invoked by malware samples. Syscall records the Linux system calls by IoT malware. The Unique Event column specifies the number of unique call used among all sequences in the dataset. The Classes column specifies the number of malware family labels in the dataset.

\begin{table*}[htbp]
\caption{Table of datasets}
\begin{center}
\begin{tabular}{|c|c|c|c|c|c|}
\hline
\textbf{Dataset} & 
\textbf{Sequence Type} & 
\textbf{Dataset Size} &
\textbf{Seq. Length} &
\textbf{Unique Event}&
\textbf{Classes}\\
\hline
SynData & 
Synthesized Alphabet Sequence with Noise&
360 &
50 &
64 &
3\\
\hline
RanSyn & 
Random Generated Synthesized Alphabet Sequence with Noise &
360 &
50 &
64 &
3\\
\hline
RanMarkov & 
Random Walk on Synthesized Win32 API Call Markov Chain&
400 &
250 &
5 &
4\\
\hline
WinMal \cite{MalWebsite}&
Win32 API Call Sequence \cite{Senpai}&
1940 &
275$^{\mathrm{a}}$& % Use average length
26&
14\\
\hline
Syscall \cite{MalWebsite}&
Linux System Call Sequence&
1208 &
364$^{\mathrm{a}}$&
121&
4\\
\hline
Oliveira\cite{IEEEdata}&
Win32 API Call Sequence from Cuckoo\cite{Cuckoo}&
3237&
100 &
307&
2\\
\hline
\multicolumn{1}{l}{$^{\mathrm{a}}$In average.}
\end{tabular}
\label{dataset}
\end{center}
\vspace{-3mm} % reduce blank between table and text
\end{table*}

\subsubsection{SynData}
There are 300 different call sequences in 3 families. The length of each call sequence is 50. The purpose of using synthetic data is to prove the correctness of our attention mechanism. Therefore, we randomly generate 50 calls which is the number between 0 to 50. After that, we replace the specific digit (21st to 25th) with the alphabet. As shown in Fig.~\ref{fig:syncalleg}, in family 1, we replace the digit with lowercase \textbf{a, b, c}, and \textbf{d} at 21st-24th digits; in family 2, replace it with \textbf{e, f, g}, and \textbf{h} 21st-24th digits; and in family 3, replace it with \textbf{i, j, k}, and \textbf{l} 21st-24th digits. Last, the 25th digit in each family is the capital letter \textbf{A}. We simulate the ordinary call with random digit and give each family a specific feature. The capital letter \text{A} is the same feature across the family. Therefore, we expect our model should only point out the lowercase alphabets are more important after training.

\begin{figure}[htbp]
    % \centering
    % \includegraphics{}
    \begin{lstlisting}[label={syncalleg}, style=myStyle,linewidth=0.48\textwidth]
Family 1: ['11','8', 'a', 'b', 'c', 'd', 'A',
'35','42','28, '13', ... ,'25','48','9']
Family 2: ['14','8', 'e', 'f', 'g', 'h', 'A',
'12','35','21','18', ... ,'28','45','1']
Family 3: ['16','4', 'i', 'j', 'k', 'l', 'A',
'17','33','29','12', ... ,'28','45','1']
    \end{lstlisting}
    \caption{Example of our synthetic API call sequence data}
    \label{fig:syncalleg}
\end{figure}

\begin{comment}
\begin{lstlisting}[label={syncalleg}, style=myStyle, caption=Example of our synthetic API call sequence data]
Family 1: ['46', '8', 'a', 'b', 'c', 'd', 'A', '35']
Family 2: ['16', '4', 'e', 'f', 'g', 'h', 'A', '12']
Family 3: ['33', '2', 'i', 'j', 'k', 'l', 'A', '21']
\end{lstlisting}
\end{comment}

\subsubsection{RanSyn}
To imitate the invocation of calls in real-world situations, we create another dataset -- RanSyn (Random generated Synthetic data). This dataset is more sophisticated than Syndata. In this dataset, the position of the alphabet subsequence is not fixed, and the numeric digit is possibly inserted between the alphabet subsequence. There are 10 percent The next digit of the numeric digit has a 10 percent probability of being an alphabet digit and 90 percent of being a numeric one. On the other hand, the next digit of the alphabet digit has a 5 percent probability of being a numeric digit. In addition, the alphabet subsequence may invoke not only one time. The other setting is the same as Syndata. 

\subsubsection{RanMarkov}
\begin{figure}[htbp]
\centerline{\includegraphics[trim = {2mm 2mm 2mm 3mm}, clip, width=0.48\textwidth]{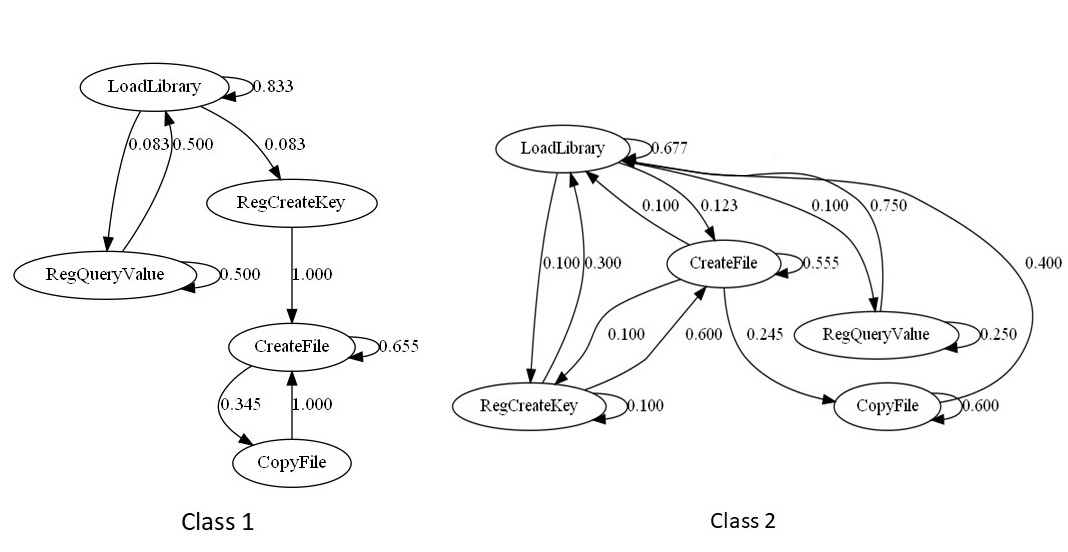}}
\caption{The synthesized Markov chain in RanMarkov dataset.}
\label{fig:markov_data}
\end{figure}

There are 400 different call sequences in 4 families. We build 4 different Markov chain at first and use random walk to traverse them to generate the call sequences whose length is 250. For example, we use five same API calls but with different transition probability and structures in Figure~\ref{fig:markov_data}. By this approach, the call sequences will have the characteristic of the Markov chain and the call sequences can be similar to the real world call sequences data.

\subsubsection{WinMal}
\begin{figure}[htbp]
\centerline{\includegraphics[trim = {5mm 8mm 3mm 3mm}, clip, width=0.48\textwidth]{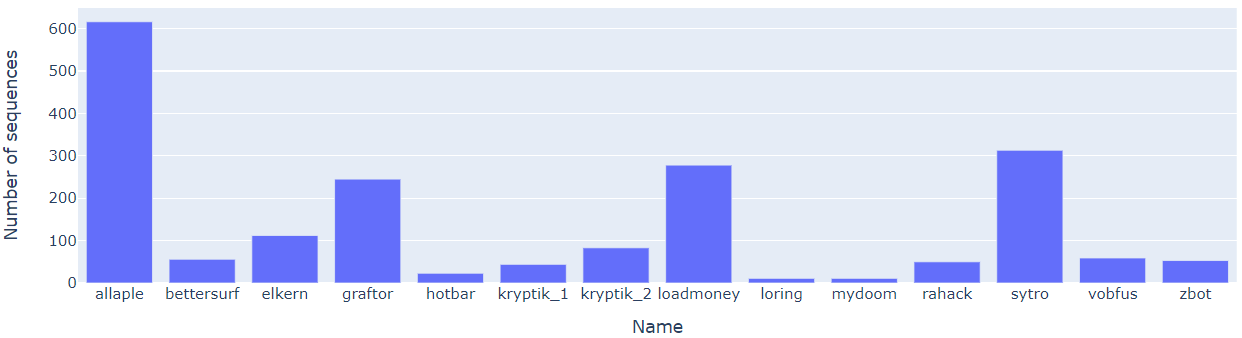}}
\caption{The API call sequences family distribution in WinMal}
\label{fig:num_bar}
\end{figure}

There are 1,940 different call sequences in 14 families. The maximum length of our API call sequences is 20,365, minimum length is 15, and average length as shown in Table~\ref{dataset}. Figure.~\ref{fig:num_bar} shows the number of API call sequences in each malware family. The malware family contains the most samples is Allaple which has 615 different profiles. In contrast, Loring and Mydoom contain the least malware samples in WinMal. On the average, the malware samples in the other families have 22-312 profiles in WinMal. Because WinMal is the real-world dataset, the imbalanced number of sample in each family is natural. Therefore, we did not adjust the number of sample to be balance. We want to test the capability of AWGCN in real-world imbalanced data.

Figure.~\ref{fig:APIcalleg} is a partial of real malware sample of API call sequence and parameters. Each file has the process identification in the beginning of profiles, and the number with pound sign is the timestamp. The content between two timestamps is the API call and its parameters.
From this sample we observe several different calls, including LoadLibrary, RegQueryValue, CreateFile, RegEnumValue. Each call has parameters to record the event state, for example, LoadLibrary records the loaded dynamic link library, CreateFile records the file stored path.
The malware samples in this dataset are available on~\cite{MalWebsite} and similar to~\cite{Senpai}. 

% 補多一點多call 多樣
\begin{figure}[htbp]
    % \centering
    % \includegraphics{}
    \begin{lstlisting}[label={APIcalleg}, style=myStyle, linewidth=0.48\textwidth]
3152 malware.exe
#209990000
LoadLibrary
lpFileName=C:\WINDOWS\system32\IMM32.DLL
Return=SUCCESS
#216910000
RegCreateKey
hKey=HKEY_CLASSES_ROOT\CLSID\
{4AEDBC33-8B19-7F8D-B932-E844B2219184}
Return=0
#216960000
RegSetValue
hKey=HKEY_CLASSES_ROOT\CLSID\
{4AEDBC33-8B19-7F8D-B932-E844B2219184}
type=REG_SZ
data=erbkznenwqlwkknq
Return=0
#216990000
RegCreateKey
hKey=HKEY_CLASSES_ROOT\CLSID\
{4AEDBC33-8B19-7F8D-B932-
E844B2219184}\LocalServer32
Return=0
#217010000
RegSetValue
hKey=HKEY_CLASSES_ROOT\CLSID\
{4AEDBC33-8B19-7F8D-B932-
E844B2219184}\LocalServer32
type=REG_SZ
data=C:\Documents and Settings\All Users\
Desktop\malware.exe
Return=0
#217260000
RegQueryValue
hKey=HKEY_LOCAL_MACHINE\System\
CurrentControlSet\Services\WinSock2\Parameters\
WinSock_Registry_Version
Return=0
type=REG_SZ
data=2.0
#217830000
CreateProcessInternal
lpCommandLine=C:\WINDOWS\system32\urdvxc.exe/
installservice
Return=SUCCESS
dwProcessId=3172
dwThreadId=3176
    \end{lstlisting}
    \caption{A partial Example of Windows API call sequence}
    \label{fig:APIcalleg}
\end{figure}

\subsubsection{Syscall}
The sequences in Syscall are longer than sequences in WinMal, and the number of unique calls is five times more than WinMal as well. Syscall contains Linux system call 1,208 sequences equally divided into four different families that collected from IoT devices. Furthermore, some English words are used in system call, such as access, open, close, etc. Unlike the call name used in Windows, combine the words directly, Linux system call has some daily used word that may be identify by the language models. This dataset is also available on~\cite{MalWebsite}.

\subsubsection{Oliveira}
The second Windows API call dataset is an open access dataset from IEEE dataport~\cite{IEEEdata}. This dataset contains 42,797 malware API call sequences and 1,079 goodware API call sequences. In our experiment, to avoid the baseline model only predicting the sample sequences as malware label, we balanced the distribution of labels. We randomly choose 2,158 malware sample, two times as goodware, the balance data can help us to confirm the correctness of the classification indicators. 
Each API call sequence is composed of the first 100 non-repeated consecutive API calls associated with the parent process. Author extract from the 'calls' elements of Cuckoo Sandbox report~\cite{Cuckoo}. Each API call sequence has its hash name, API call name (in integer type), and the label with 0 for goodware and 1 for malware. Therefore, we can use binary classification for downstream task to test our proposed model.

\subsection{Family Classification Camparsion}
In this experiment, we compare our Attention Aware GCN (AWGCN) with text-based methods and the different classifier shown in Table~\ref{evaluation} on four datasets. The number in first row is test accuracy, second row is f1-score, and third row is area under curve (AUC) score.

Because the original synthetic data is too simple to predict (each family contain specific feature), we increase the classification difficulty via making the sequences more complicated by adding noise.
We create noise sequences by adding some noise calls into the original synthetic sequences. We add the other two families specific features into the call sequences in the family. For example, we add \textbf{e, f, g, h, i, j, k}, and \textbf{l} into family 1, and so on. Moreover, we add a random integer call between the original feature of the family, such as \textbf{a, b, 5, c}, and \textbf{d}. Namely, each sequence contains the all the original family features.
In addition, the training data in this Syndata and RanSyn are synthetic data without noise, we use 60 noise sequences only in testing data to test the generalization ability of the models. 

\begin{figure}[t]
\centerline{\includegraphics[trim = {0mm 0 0 0}, clip, width=0.48\textwidth]{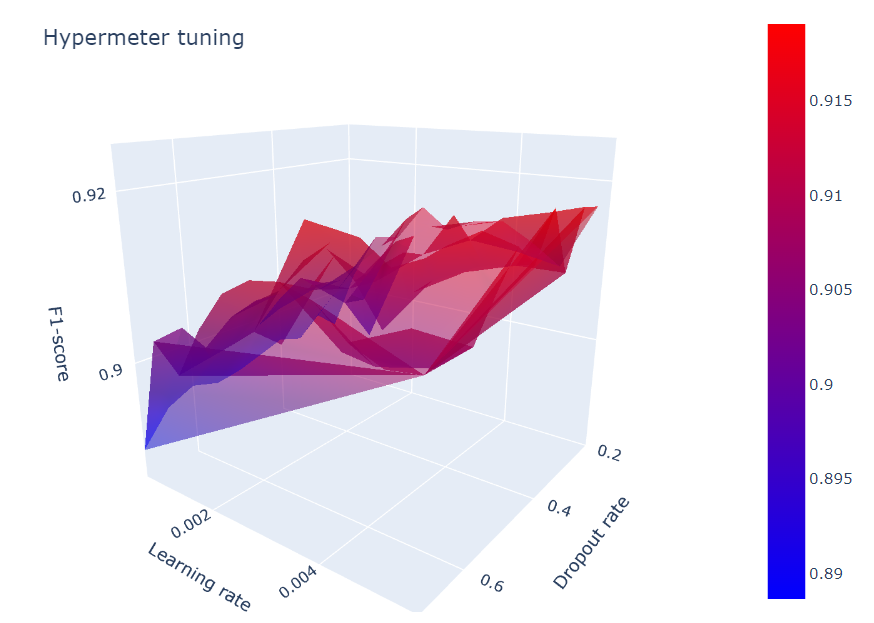}}
\caption{AWGCN hypermeter: learning and dropout rate tuning}
\label{fig:lr_dropout}
\end{figure}

% the others model setting?
For AWGCN, we use three graph convolution layers to extract the information from the graph. We also experimented with other settings to find better parameters, as shown in Figure.~\ref{fig:lr_dropout}. We can observe the lower f1-score the point color is close to blue. The peak of f1-score is close to red, and the point of the peak is close to (0.005, 0.35). Therefore, we set the learning rate as 0.005, dropout rate as 0.35, and the dimension in each graph convolution layer as \textit{$k1=128$}, \textit{$k2=256$}, and \textit{$k3=64$}, respectively. % epoch batch

\begin{figure}[t]
\centerline{\includegraphics[trim = {5mm 0 35mm 15mm}, clip, width=0.48\textwidth]{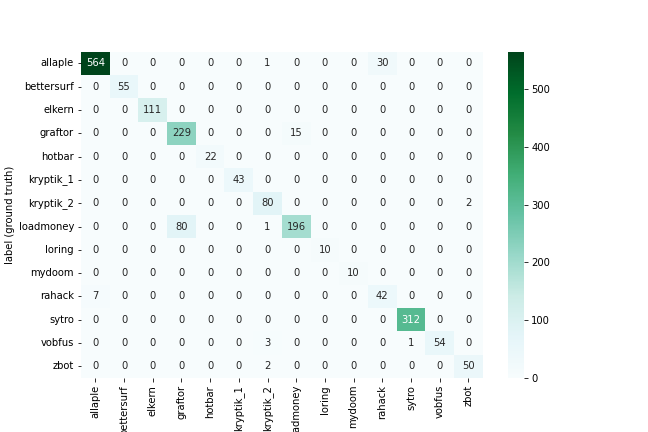}}
\caption{Classification confusion matrix of AWGCN in WinMal dataset.}
\label{fig:cf_matrix}
\end{figure}

% fig 10 獨立出來解釋 是哪個model 對角線代表啥 predict的莊況
\textbf{Performance}. Table~\ref{evaluation} presents the classification result of each model. AWGCN outperforms the other models on four different datasets, especially on real-world malware dataset. Figure.~\ref{fig:cf_matrix} show the confusion matrix of the classification result on WinMal dataset. The diagonal line represent the correct predictions of the model. The families contain least sample, Loring and Mydoom, AWGCN can perfectly classify them into correct malware family. The most misclassification of AWGCN is between the family Loadmoney and Graftor. According to introduce of these two malware family of the anti-virus company Malwarebytes~\cite{Loadmoney, Graftor}, the malware Loadmoney are known to install adware, browser hijackers, and potentially unwanted programs (PUPs) on Windows systems. The malware Graftor is a large family of malware targeting Windows systems, most of which are Trojans, though some are adware. These two family have prefix word adware on the citation website. AWGCN classify the malware samples via the behavior of malware, the misclassification may be the similar action but in two different families.

AWGCN is better than others because AWGCN investigates the transition information from the original call sequences. For in-depth analysis, sequence alignment can perform well if the features are in a specific position. If some unique calls are only used in a specific family, the primary classifier with one-hot encoding can be a fast approach with acceptable accuracy. Text-based approaches also performed nice results, but they are more suitable for language corpus than heavily repetitive sequential data like API call sequence.

In Table~\ref{lat_comparison}, we use the same model to identify whether AWGCN extracts the graph representation successfully or not. We output the graph latent space vector (shown as blue vector in Fig.\ref{fig:Ch3_fig}) to compare with the one-hot encoding data. As the result show in Table~\ref{lat_comparison}, AWGCN latent space improve the performance of classical models. In the experiment of SVM, AWGCN enhance performance around 14 percent. This experiment indicate that AWGCN can perform a well embedding of original sequences and achieves significant improvement in extracting the representation.

\begin{table*}[t]
\caption{Results of family classification task.}
\begin{center}
\begin{tabular}{ccccccccc}
\hline
\textbf{}&
\textbf{Encoding} & 
\textbf{Model} &
\textbf{SynData} &
\textbf{RanSyn}&
\textbf{RanMarkov}&
\textbf{WinMal}&
\textbf{Syscall}&
\textbf{Oliveira}\\
\hline
\multirow{3}{*}{SeqAm}&
\multirow{3}{*}{-}&
\multirow{3}{*}{Seq. Alignment} &  0.987&0.300&0.900&0.759&0.401&0.485\\
                                &&& 0.994&0.821&0.900&0.759&0.401&0.485\\ 
                                &&& 0.996&0.942&0.973&0.963&0.755&0.571\\
\hline
\multirow{3}{*}{OHSVM}&
\multirow{3}{*}{One-hot}&
\multirow{3}{*}{SVM}&   0.317&0.487&0.738&0.812&0.945&0.739\\ 
                    &&& 0.300&0.351&0.703&0.791&0.945&0.741\\ 
                    &&& 0.485&0.693&0.910&0.973&0.988&0.811\\
\hline
\multirow{3}{*}{OHLR}&
\multirow{3}{*}{One-hot}&
\multirow{3}{*}{Logistic Regression}&   0.333&0.463&0.738&0.855&0.948&0.864\\ 
                                    &&& 0.167&0.416&0.703&0.852&0.948&0.862\\ 
                                    &&& 0.507&0.616&0.910&0.981&0.992&0.907\\
\hline
\multirow{3}{*}{OHRF}&
\multirow{3}{*}{One-hot}&
\multirow{3}{*}{Random Forest}&     0.367&0.487&0.738&0.873&0.948&0.904\\ 
                              &&&   0.278&0.447&0.703&0.871&0.948&0.904\\
                              &&&   0.512&0.619&0.910&0.982&0.992&0.970\\

\hline
\multirow{3}{*}{Engtk-BiLSTM}&
\multirow{3}{*}{spaCy tokenizer}&
\multirow[c]{2}{*}{Bi-LSTM}&\textbf{1.000}&0.742&\textbf{1.000}&0.914&0.964&0.941\\
&&\multirow[b]{2}{*}{Dense}&\textbf{1.000}&0.945&\textbf{1.000}&0.906&0.965&0.974\\ 
                            &&&\textbf{1.000}&0.988&\textbf{1.000}&0.992&0.995&0.973\\

\hline
\multirow{3}{*}{T2V-BiLSTM~\cite{tensorflow-rnn}}&
\multirow{3}{*}{TextVectorization}&
\multirow[c]{2}{*}{Bi-LSTM}&0.275&0.250&0.996&0.882&0.930&0.922\\ 
&&\multirow[b]{2}{*}{Dense}&0.311&0.338&0.997&0.833&0.949&0.959\\ 
                         &&&0.485&0.506&0.999&0.972&0.990&0.955\\
\hline
\multirow{3}{*}{W2V-GRU}&
\multirow{3}{*}{Word2Vec}&
\multirow[c]{2}{*}{GRU}     &0.250&0.237&\textbf{1.000}&0.914&0.933&0.837\\ 
&&\multirow[b]{2}{*}{Dense} &0.333&0.830&\textbf{1.000}&0.886&0.923&0.843\\ 
                            &&& 0.493&0.972&\textbf{1.000}&0.989&0.990&0.819\\
\hline
\multirow{3}{*}{TkCNN}&
\multirow{3}{*}{Tokenizer}&
\multirow[c]{2}{*}{1D CNN}  &\textbf{1.000}&0.475&0.986&0.919&0.958&0.933\\ 
&&\multirow[b]{2}{*}{Dense} &\textbf{1.000}&0.886&0.994&0.905&0.962&0.969\\ 
                            &&&\textbf{1.000}&0.917&0.999&0.990&0.995&0.965\\
\hline
\multirow{3}{*}{Bert~\cite{tensorflow-bert}}&
\multirow{3}{*}{Bert}&
\multirow{3}{*}{Dense}  &0.500&0.524&0.850&0.911&0.935&0.945\\ 
                        &&&0.888&0.894&0.937&0.922&0.950&0.977\\ 
                        &&&0.995&0.993&0.989&0.983&0.991&0.974\\
\hline
\multirow{3}{*}{AWGCN}&
\multirow[c]{2}{*}{One-hot}&
\multirow[c]{2}{*}{AWGCN}&\textbf{1.000}&\textbf{0.912}&\textbf{1.000}&\textbf{0.927}&\textbf{0.966}&\textbf{0.992}\\& 
\multirow[b]{2}{*}{Adjacecny}&\multirow[b]{2}{*}{Dense}&\textbf{1.000}&\textbf{0.980}&\textbf{1.000}&\textbf{0.941}&\textbf{0.966}&\textbf{0.998}\\ 
&&&\textbf{1.000}&\textbf{0.998}&\textbf{1.000}&\textbf{0.995}&\textbf{0.996}&\textbf{0.997}\\
\hline
\multicolumn{7}{l}{$*$The numbers in each row represent Testing Accuracy, F1-score, and AUC-score, respectively.}
\end{tabular}
\label{evaluation}
\end{center}
\vspace{-5mm} % reduce blank between table and text
\end{table*}

\begin{table}[t]
\setlength{\tabcolsep}{2pt}
\caption{Comparison between one-hot and AWGCN latent space}
\begin{center}
\begin{tabular}{ccc}
\hline
\textbf{Encoding/Embedding} & \textbf{Model} &\textbf{Indicators}\\
\hline
\multirow{9}{*}{One-hot}&
\multirow{3}{*}{SVM}&0.812\\
                    &&0.791\\
                    &&0.973\\
\cline{2-3}
& \multirow{3}{*}{Logistic regression}&0.855\\
                                      &&0.852\\
                                      &&0.981\\
\cline{2-3}
& \multirow{3}{*}{Random Forest}&0.873\\
                                &&0.871\\
                                &&0.982\\
\hline
\multirow{9}{*}{AWGCN latent space}&
\multirow{3}{*}{SVM}&\textbf{0.926}\\
                    &&\textbf{0.927}\\
                    &&\textbf{0.992}\\
\cline{2-3}
& \multirow{3}{*}{Logistic regression}&\textbf{0.923}\\
                                      &&\textbf{0.924}\\
                                      &&\textbf{0.993}\\
\cline{2-3}
& \multirow{3}{*}{Random Forest}&\textbf{0.912}\\
                                &&\textbf{0.912}\\
                                &&\textbf{0.987}\\
\hline
\end{tabular}
\label{lat_comparison}
\end{center}
\vspace{-3mm} % reduce blank between table and text
\end{table}

\subsection{Attention Mechanism}
In this section, we evaluate our designed attention mechanism via attention visualization.

\begin{figure*}[t]
\centerline{\includegraphics[trim = {0mm 0 0 0}, clip, width=\textwidth]{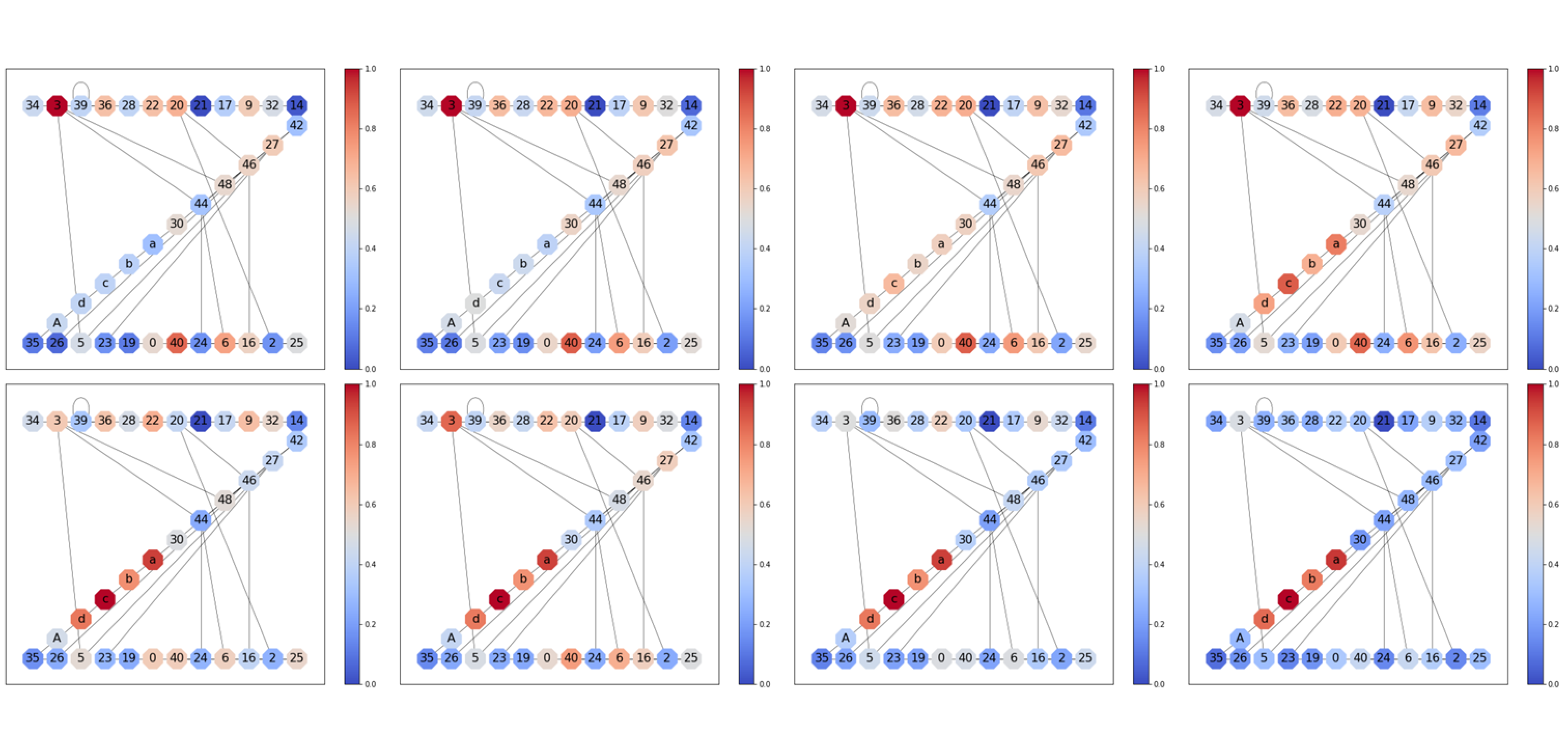}}
\caption{The example of the attention transition of synthetic API call sequence.}
\label{fig:syn_att_trans}
\end{figure*}

As mentioned above, synthetic sequences are constructed by ordinary calls (random numbers) and specific calls (alphabet sequences). In other words, our designed mechanism should only point out the alphabet sub-sequence.
In Fig.~\ref{fig:syn_att_trans}, we visualize the attention weights on synthetic data. The upper left is the beginning epoch of our training process, and the right-hand side is the next epoch. We can easily observe that each node's attention weight changes when training. The weight of first epoch is random initialized with uniform distribution; thus, the alphabet sequence seems to be insignificant, and the attention hot spot is on ordinary calls. However, by the training process moving on, the feature attention hot spot are gradually transfer to the feature of the sequence. After the whole training process, the attention weights on the bottom right are focused on the specific features of the family (alphabet sequence). Namely, the attention mechanism of AWGCN is trainable and can precisely point out the significant part of the graph after training.

\begin{figure}[t]
\centerline{\includegraphics[trim = {0mm 0mm 0mm 0mm}, clip, width=0.48\textwidth]{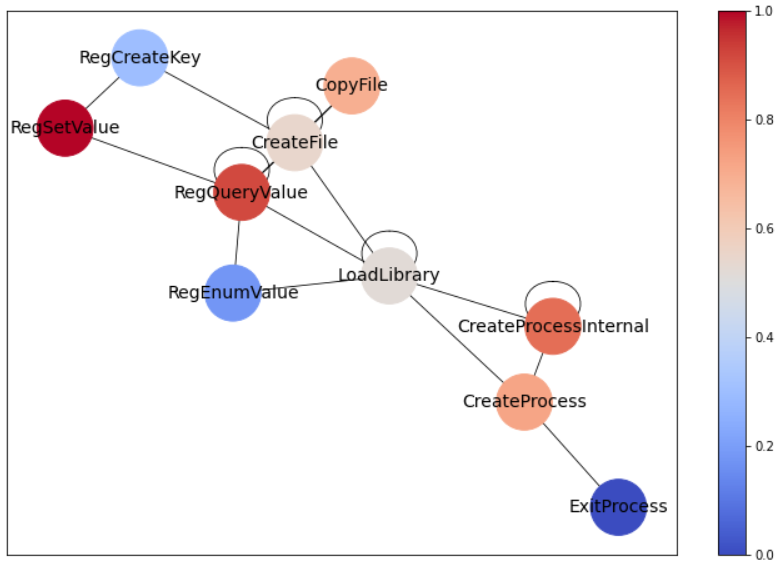}}
\caption{The attention result of real world malware sample in WinMal dataset.}
\label{fig:real_sample}
\end{figure}

\begin{figure}[t]
\centerline{\includegraphics[trim = {0mm 12mm 0mm 12mm}, clip, width=0.48\textwidth]{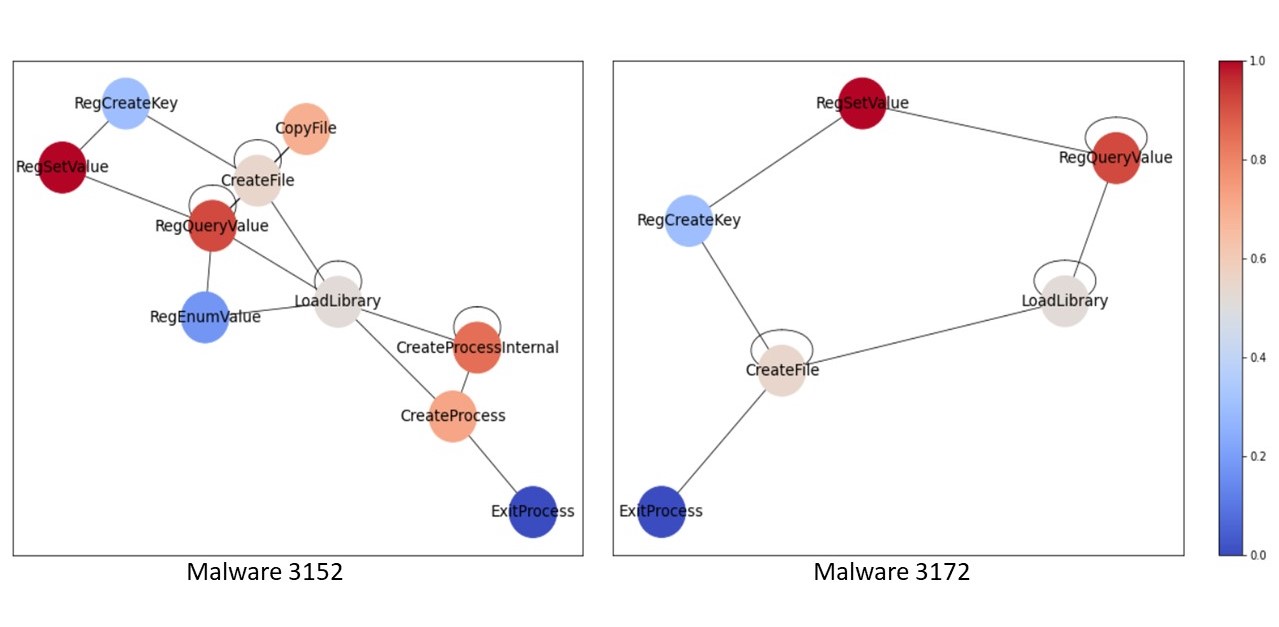}}
\caption{The sample of multi-threaded process of Allaple malware in WinMal dataset.}
\label{fig:sample_camparsion}
\end{figure}

In Figure.~\ref{fig:real_sample}, we demonstrate the real world malware example in WinMal dataset. The result shows that RegSetValue, RegQueryValue, and CreateProcessInternal are more critical in this malware sample. According to the report~\cite{Trendmicro, Microsoft}, this malware is a worm and named Allaple. The report said that Allaple modifies the registry to reference a unique CLSID, in our original malware profile, this malware call the RegSetValue after create the registry key and activate the malware.exe stored at Desktop. Furthermore, according to the second report from Microsoft, this malware is a multi-threaded worm and it copies itself to the Windows system folder using the filename "urdvxc.exe" or "irdvxc.exe" and modifies the registry to load this copy when Windows is started. In our profiles, RegQueryValue query the path mentioned in report to CreateProcessInternal using the filename "urdvxc.exe" to create another process, as shown in Figure.~\ref{fig:APIcalleg}. Moreover, the call CreateFile calls the rsaenh.dll to encrypt at last which is pointed out by our attention mechanism. In summary, attention mechanism can properly figure out the critical part of the input API call sequence and the result is similar to the report provided by antivirus companies.

In Figure.~\ref{fig:sample_camparsion}, we show the created multi-threaded process from malware 3152. They have same hash value with different suffix identification number and they similar behavior as well. The difference between malware 3152 and 3172 is that malware 3152 call CreateProcess to create another process for its following behavior. 

\subsection{Representation of Malware Family}
In this part, we visualize the graph latent space training by AWGCN of the WinMal dataset. Because WinMal dataset has 1,940 samples and is distributed in 14 families, which can better represent the graph latent space, we use WinMal dataset as an example to visualize the result.

\begin{figure}[t]
\centerline{\includegraphics[trim = {0mm 0 0 0}, clip, width=0.48\textwidth]{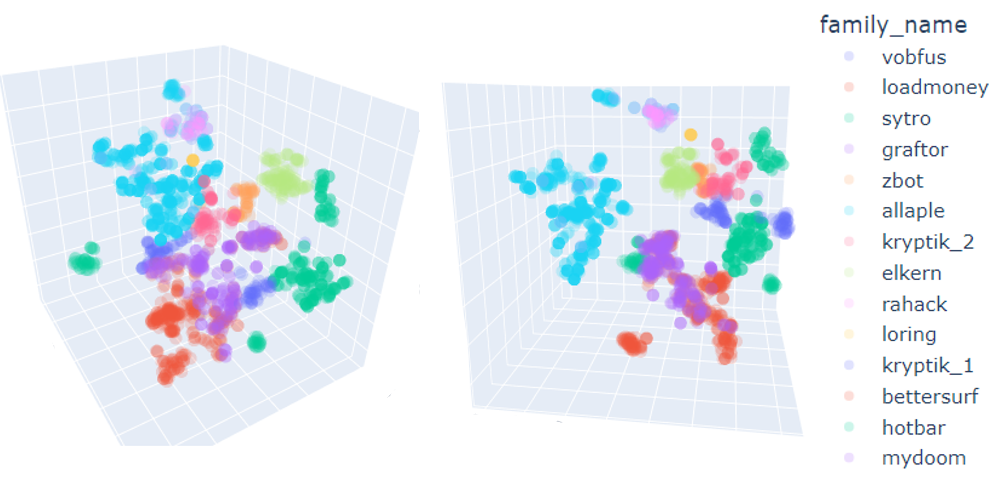}}
\caption{The malware family distribution in latent space of WinMal dataset.}
\label{fig:tSNE}
\end{figure}

In Figure.~\ref{fig:tSNE}, we use t-SNE to do dimension reduction in order to visualize. The t-SNE result in Figure.~\ref{fig:tSNE} is the same space but in different angle. (You can access the auto rotation and interaction t-SNE on this link, \url{https://github.com/ChuPoYu/14family_latent_space}). In Figure.~\ref{fig:tSNE}, we can observe that most of the malware family are been separated to different area in t-SNE. Only the partial samples in Loadmoney and Graftor are overlapped (the purple and red dot in right part of Fig.~\ref{fig:tSNE} at the bottom). The main reason is same as mentioned above, there are some samples with similar action but in different family label.

\begin{figure}[t]
\centerline{\includegraphics[trim = {0mm 0 0 0}, clip, width=0.48\textwidth]{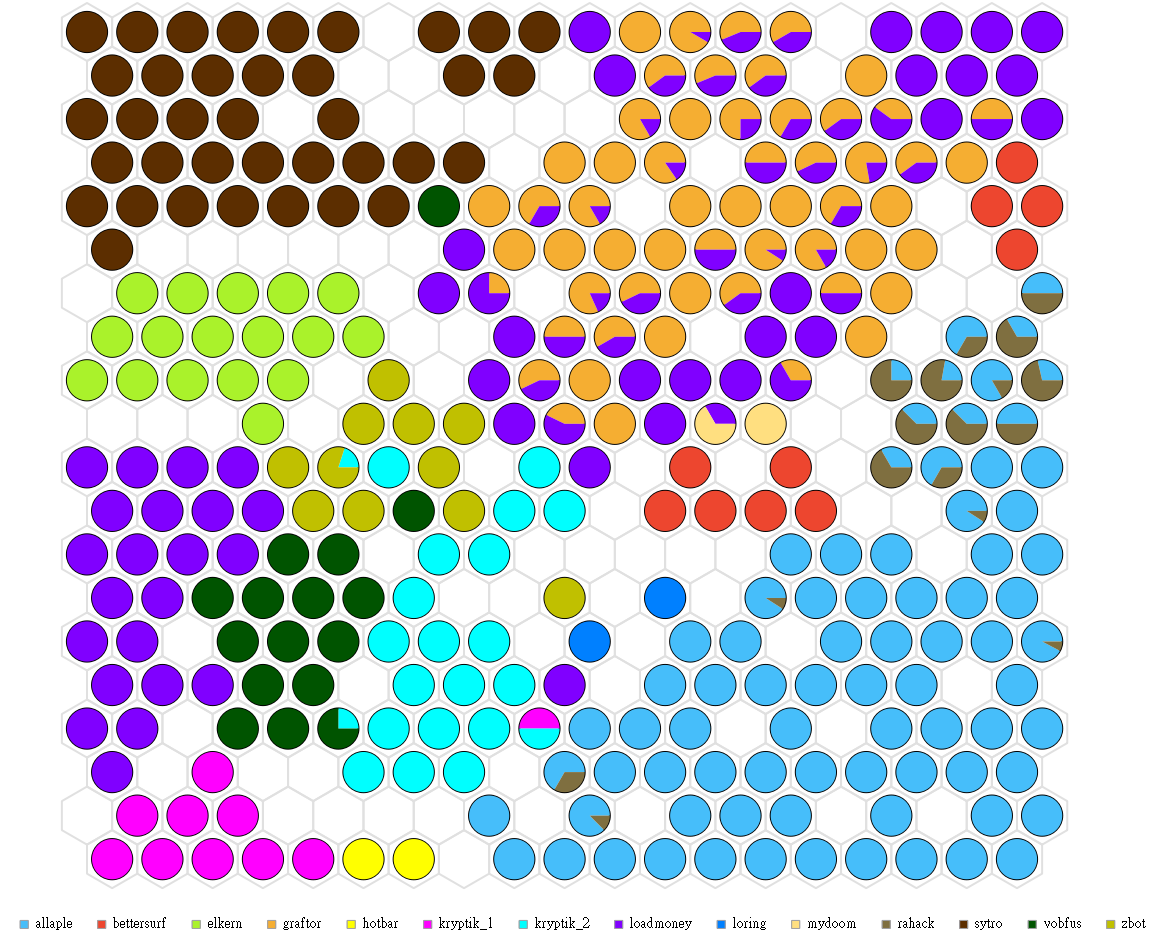}}
\caption{The SOM clustering with malware family label.}
\label{fig:SOM_family}
\end{figure}

In Figure.~\ref{fig:SOM_family}, we perform a self organizing map (SOM) clustering on our graph latent space vector. Most of the malware profile are clustered to a proper space, for instance, the samples in Allaple family (the dot colored in ocean blue at the bottom-right in Fig.~\ref{fig:SOM_family}). This result show that the samples in Allaple have similar action; thus, they are clustered to closed area. In contrast to Allaple, the samples in Loadmoney family (the dot colored purple) are clustered to two different area. This indicate Loadmoeny family may have two branches, one have similar action to Grafter, another one do different actions.

\begin{figure}[t]
\centerline{\includegraphics[trim = {20mm 0 65mm 0mm}, clip, width=0.48\textwidth]{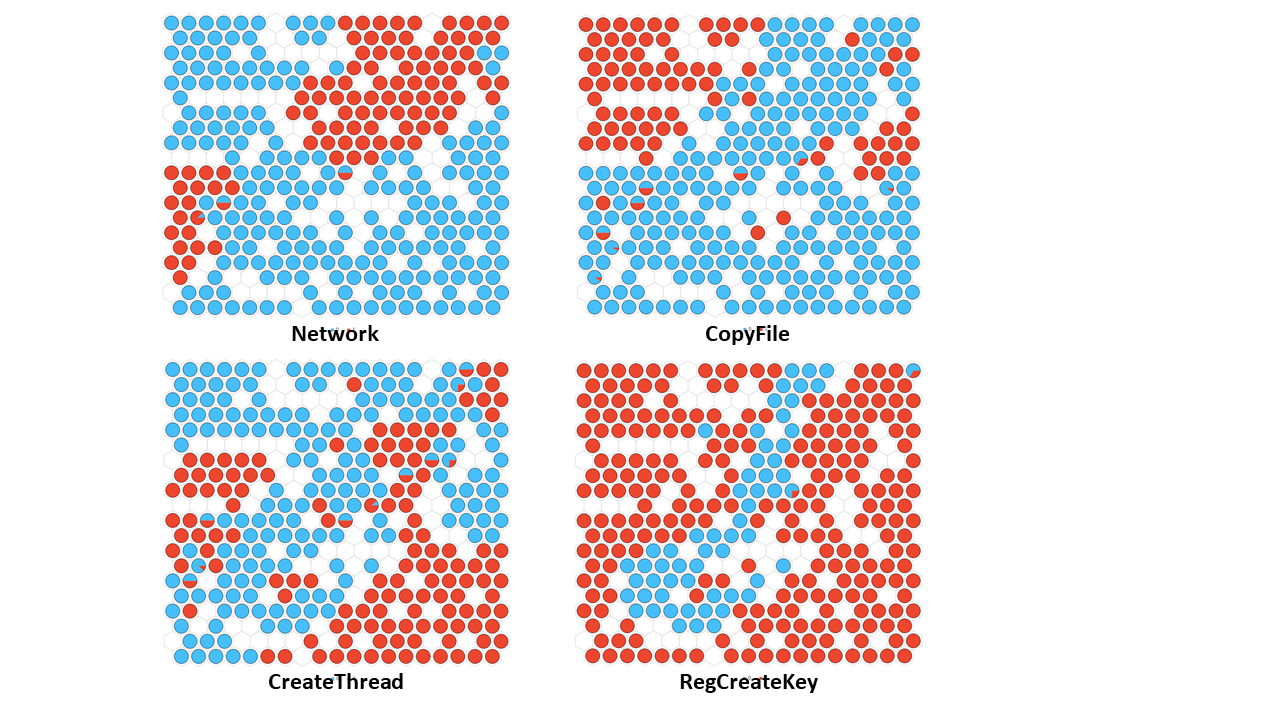}}
\caption{The SOM containing specific call.}
\label{fig:SOM}
\end{figure}

Because we want to know will the used calls affect the malware family distribution. In Figure.~\ref{fig:SOM}, we use different four keys to visualize the same SOM clustering, including CopyFile, CreateThread, RegCreateKey, and Network related calls (e.g. WinHttpConnect and InternetOpen). In Figure.~\ref{fig:SOM}, the malware use network mainly separate to two area, and they are belong to LoadMoney and Graftor. The malware use CopyFile are mainly belong to Sytro, which is a Worm family and can copy and spread themselves. Through Figure.~\ref{fig:SOM_family} and Figure.~\ref{fig:SOM} , we can realize what may be the main action of the malware and which family are they belong to.

\section{Discussion}\label{sec:discussion}
The experiment results show that the proposed AWGCN achieves strong family classification results and learns the graph embedding in a better representation. Moreover, AWGCN can make the inductive prediction for unseen sequence samples. 

In the future, we can test AWGCN on different datasets but have similar properties, such as event log data which is highly repetitive. Furthermore, we can replace Markov model with other graph representation to further improve the information preserving from original sequences. Last but not least, we can perform AWGCN back to the text data to test the generalization and the correctness of attention in the natural language processing domain.

\section{Conclusion}\label{sec:conclusion}
In this paper, we provide a point of view from graphs to deal with structural text-based API call sequential data. The classification experiment results show that using our designed AWGCN to analyze the call-like sequential data outperforms other methods and properly separates the data points in the latent space. Also, our method conquers the disadvantage of sequence alignment in coping with variable length sequences. Furthermore, we implement the attention mechanism, which can help us quickly point out the critical component of the malware. 

\section*{Acknowledgment}
The work is partially supported by the Ministry of Science and Technology, Taiwan under Grant No. MOST 109-2221-E-004-007-MY3 and No. MOST 111-2218-E-001-001-MBK.

% trigger a \newpage just before the given reference
% number - used to balance the columns on the last page
% adjust value as needed - may need to be readjusted if
% the document is modified later
%\IEEEtriggeratref{8}
% The "triggered" command can be changed if desired:
%\IEEEtriggercmd{\enlargethispage{-5in}}

% references section

% can use a bibliography generated by BibTeX as a .bbl file
% BibTeX documentation can be easily obtained at:
% http://www.ctan.org/tex-archive/biblio/bibtex/contrib/doc/
% The IEEEtran BibTeX style support page is at:
% http://www.michaelshell.org/tex/ieeetran/bibtex/
%\bibliographystyle{IEEEtran}
% argument is your BibTeX string definitions and bibliography database(s)
%\bibliography{IEEEabrv,../bib/paper}
%
% <OR> manually copy in the resultant .bbl file
% set second argument of \begin to the number of references
% (used to reserve space for the reference number labels box)

% that's all folks
\end{document}